\documentclass[
  superscriptaddress,
  reprint,
  amsmath,amssymb,
  aps,
  prl,
  nofootinbib,
  floatfix,
  checklength
]{revtex4-2}

\usepackage{graphicx}
\usepackage{dcolumn} 
\usepackage{bm}
\usepackage[table]{xcolor}
\usepackage{colortbl}
\usepackage{float}
\usepackage{afterpage}
\usepackage{microtype}
\usepackage[linesnumbered,ruled]{algorithm2e}
\usepackage{url}
\usepackage{multirow,makecell}
\usepackage{booktabs}
\usepackage[caption=false]{subfig}

\usepackage{hyperref}
\hypersetup{
  hidelinks,
  colorlinks=true,
  citecolor=teal,
  linkcolor=teal,
  urlcolor=teal
}

\usepackage[colorinlistoftodos]{todonotes}

\usepackage{lipsum}

\usepackage{parskip}
\begin{document}

\title{First double-differential measurement of pionless charged-current muon neutrino interactions using kinematic imbalance observables on carbon and oxygen with the T2K experiment}

\newcommand{\INSTHD}{\affiliation{University Autonoma Madrid, Department of Theoretical Physics, 28049 Madrid, Spain}}
\newcommand{\INSTFE}{\affiliation{Boston University, Department of Physics, Boston, Massachusetts, U.S.A.}}
\newcommand{\INSTD}{\affiliation{University of British Columbia, Department of Physics and Astronomy, Vancouver, British Columbia, Canada}}
\newcommand{\INSTGA}{\affiliation{University of California, Irvine, Department of Physics and Astronomy, Irvine, California, U.S.A.}}
\newcommand{\INSTI}{\affiliation{IRFU, CEA, Universit\'e Paris-Saclay, F-91191 Gif-sur-Yvette, France}}
\newcommand{\INSTGB}{\affiliation{University of Colorado at Boulder, Department of Physics, Boulder, Colorado, U.S.A.}}
\newcommand{\INSTFH}{\affiliation{Duke University, Department of Physics, Durham, North Carolina, U.S.A.}}
\newcommand{\INSTEF}{\affiliation{ETH Zurich, Institute for Particle Physics and Astrophysics, Zurich, Switzerland}}
\newcommand{\INSTIG}{\affiliation{VNU University of Science, Vietnam National University, Hanoi, Vietnam}}
\newcommand{\INSTIE}{\affiliation{CERN European Organization for Nuclear Research, CH-1211 Gen\'eve 23, Switzerland}}
\newcommand{\INSTEG}{\affiliation{University of Geneva, Section de Physique, DPNC, Geneva, Switzerland}}
\newcommand{\INSTHJ}{\affiliation{University of Glasgow, School of Physics and Astronomy, Glasgow, United Kingdom}}
\newcommand{\INSTJG}{\affiliation{Ghent University, Department of Physics and Astronomy, Proeftuinstraat 86, B-9000 Gent, Belgium}}
\newcommand{\INSTDG}{\affiliation{H. Niewodniczanski Institute of Nuclear Physics PAN, Cracow, Poland}}
\newcommand{\INSTCB}{\affiliation{High Energy Accelerator Research Organization (KEK), Tsukuba, Ibaraki, Japan}}
\newcommand{\INSTIB}{\affiliation{University of Houston, Department of Physics, Houston, Texas, U.S.A.}}
\newcommand{\INSTED}{\affiliation{Institut de Fisica d'Altes Energies (IFAE) - The Barcelona Institute of Science and Technology, Campus UAB, Bellaterra (Barcelona) Spain}}
\newcommand{\INSTJC}{\affiliation{Institut f\"ur Physik, Johannes Gutenberg-Universit\"at Mainz, Staudingerweg 7, 55128 Mainz, Germany}}
\newcommand{\INSTHH}{\affiliation{Institute For Interdisciplinary Research in Science and Education (IFIRSE), ICISE, Quy Nhon, Vietnam}}
\newcommand{\INSTEI}{\affiliation{Imperial College London, Department of Physics, London, United Kingdom}}
\newcommand{\INSTGF}{\affiliation{INFN Sezione di Bari and Universit\`a e Politecnico di Bari, Dipartimento Interuniversitario di Fisica, Bari, Italy}}
\newcommand{\INSTBE}{\affiliation{INFN Sezione di Napoli and Universit\`a di Napoli, Dipartimento di Fisica, Napoli, Italy}}
\newcommand{\INSTBF}{\affiliation{INFN Sezione di Padova and Universit\`a di Padova, Dipartimento di Fisica, Padova, Italy}}
\newcommand{\INSTBD}{\affiliation{INFN Sezione di Roma and Universit\`a di Roma ``La Sapienza'', Roma, Italy}}
\newcommand{\INSTEB}{\affiliation{Institute for Nuclear Research of the Russian Academy of Sciences, Moscow, Russia}}
\newcommand{\INSTHI}{\affiliation{International Centre of Physics, Institute of Physics (IOP), Vietnam Academy of Science and Technology (VAST), 10 Dao Tan, Ba Dinh, Hanoi, Vietnam}}
\newcommand{\INSTJD}{\affiliation{ILANCE, CNRS – University of Tokyo International Research Laboratory, Kashiwa, Chiba 277-8582, Japan}}
\newcommand{\INSTHA}{\affiliation{Kavli Institute for the Physics and Mathematics of the Universe (WPI), The University of Tokyo Institutes for Advanced Study, University of Tokyo, Kashiwa, Chiba, Japan}}
\newcommand{\INSTID}{\affiliation{Keio University, Department of Physics, Kanagawa, Japan}}
\newcommand{\INSTIF}{\affiliation{King's College London, Department of Physics, Strand, London WC2R 2LS, United Kingdom}}
\newcommand{\INSTCC}{\affiliation{Kobe University, Kobe, Japan}}
\newcommand{\INSTCD}{\affiliation{Kyoto University, Department of Physics, Kyoto, Japan}}
\newcommand{\INSTEJ}{\affiliation{Lancaster University, School of Physics and Astronomy, Lancaster, United Kingdom}}
\newcommand{\INSTII}{\affiliation{Lawrence Berkeley National Laboratory, Berkeley, California, U.S.A.}}
\newcommand{\INSTBA}{\affiliation{Ecole Polytechnique, IN2P3-CNRS, Laboratoire Leprince-Ringuet, Palaiseau, France}}
\newcommand{\INSTFC}{\affiliation{University of Liverpool, Department of Physics, Liverpool, United Kingdom}}
\newcommand{\INSTFI}{\affiliation{Louisiana State University, Department of Physics and Astronomy, Baton Rouge, Louisiana, U.S.A.}}
\newcommand{\INSTIH}{\affiliation{Joint Institute for Nuclear Research, Dubna, Moscow Region, Russia}}
\newcommand{\INSTHB}{\affiliation{Michigan State University, Department of Physics and Astronomy,  East Lansing, Michigan, U.S.A.}}
\newcommand{\INSTCE}{\affiliation{Miyagi University of Education, Department of Physics, Sendai, Japan}}
\newcommand{\INSTDF}{\affiliation{National Centre for Nuclear Research, Warsaw, Poland}}
\newcommand{\INSTFJ}{\affiliation{State University of New York at Stony Brook, Department of Physics and Astronomy, Stony Brook, New York, U.S.A.}}
\newcommand{\INSTEH}{\affiliation{STFC, Rutherford Appleton Laboratory, Harwell Oxford,  and  Daresbury Laboratory, Warrington, United Kingdom}}
\newcommand{\INSTGJ}{\affiliation{Okayama University, Department of Physics, Okayama, Japan}}
\newcommand{\INSTCF}{\affiliation{Osaka Metropolitan University, Department of Physics, Osaka, Japan}}
\newcommand{\INSTGG}{\affiliation{Oxford University, Department of Physics, Oxford, United Kingdom}}
\newcommand{\INSTIC}{\affiliation{University of Pennsylvania, Department of Physics and Astronomy,  Philadelphia, Pennsylvania, U.S.A.}}
\newcommand{\INSTGC}{\affiliation{University of Pittsburgh, Department of Physics and Astronomy, Pittsburgh, Pennsylvania, U.S.A.}}
\newcommand{\INSTGD}{\affiliation{University of Rochester, Department of Physics and Astronomy, Rochester, New York, U.S.A.}}
\newcommand{\INSTHC}{\affiliation{Royal Holloway University of London, Department of Physics, Egham, Surrey, United Kingdom}}
\newcommand{\INSTBC}{\affiliation{RWTH Aachen University, III. Physikalisches Institut, Aachen, Germany}}
\newcommand{\INSTJF}{\affiliation{School of Physics and Astronomy, University of Minnesota, Minneapolis, Minnesota, U.S.A.}}
\newcommand{\INSTJB}{\affiliation{Departamento de F\'isica At\'omica, Molecular y Nuclear, Universidad de Sevilla, 41080 Sevilla, Spain}}
\newcommand{\INSTFB}{\affiliation{University of Sheffield, School of Mathematical and Physical Sciences, Sheffield, United Kingdom}}
\newcommand{\INSTDI}{\affiliation{University of Silesia, Institute of Physics, Katowice, Poland}}
\newcommand{\INSTIA}{\affiliation{SLAC National Accelerator Laboratory, Stanford University, Menlo Park, California, U.S.A.}}
\newcommand{\INSTBB}{\affiliation{Sorbonne Universit\'e, CNRS/IN2P3, Laboratoire de Physique Nucl\'eaire et de Hautes Energies (LPNHE), Paris, France}}
\newcommand{\INSTJE}{\affiliation{South Dakota School of Mines and Technology, 501 East Saint Joseph Street, Rapid City, SD 57701, United States}}
\newcommand{\INSTCH}{\affiliation{University of Tokyo, Department of Physics, Tokyo, Japan}}
\newcommand{\INSTBJ}{\affiliation{University of Tokyo, Institute for Cosmic Ray Research, Kamioka Observatory, Kamioka, Japan}}
\newcommand{\INSTCG}{\affiliation{University of Tokyo, Institute for Cosmic Ray Research, Research Center for Cosmic Neutrinos, Kashiwa, Japan}}
\newcommand{\INSTHF}{\affiliation{Institute of Science Tokyo, Department of Physics, Tokyo}}
\newcommand{\INSTGI}{\affiliation{Tokyo Metropolitan University, Department of Physics, Tokyo, Japan}}
\newcommand{\INSTHG}{\affiliation{Tokyo University of Science, Faculty of Science and Technology, Department of Physics, Noda, Chiba, Japan}}
\newcommand{\INSTB}{\affiliation{TRIUMF, Vancouver, British Columbia, Canada}}
\newcommand{\INSTJH}{\affiliation{University of Toyama, Faculty of Science, Toyama, Japan}}
\newcommand{\INSTJJ}{\affiliation{University of Tsukuba, Faculty of Pure and Applied Sciences, Tsukuba, Ibaraki, Japan}}
\newcommand{\INSTDJ}{\affiliation{University of Warsaw, Faculty of Physics, Warsaw, Poland}}
\newcommand{\INSTDH}{\affiliation{Warsaw University of Technology, Institute of Radioelectronics and Multimedia Technology, Warsaw, Poland}}
\newcommand{\INSTIJ}{\affiliation{Tohoku University, Faculty of Science, Department of Physics, Miyagi, Japan}}
\newcommand{\INSTFD}{\affiliation{University of Warwick, Department of Physics, Coventry, United Kingdom}}
\newcommand{\INSTEA}{\affiliation{Wroclaw University, Faculty of Physics and Astronomy, Wroclaw, Poland}}
\newcommand{\INSTHE}{\affiliation{Yokohama National University, Department of Physics, Yokohama, Japan}}
\newcommand{\INSTH}{\affiliation{York University, Department of Physics and Astronomy, Toronto, Ontario, Canada}}

\INSTHD
\INSTFE
\INSTD
\INSTGA
\INSTI
\INSTGB
\INSTFH
\INSTEF
\INSTIG
\INSTIE
\INSTEG
\INSTHJ
\INSTJG
\INSTDG
\INSTCB
\INSTIB
\INSTED
\INSTJC
\INSTHH
\INSTEI
\INSTGF
\INSTBE
\INSTBF
\INSTBD
\INSTEB
\INSTHI
\INSTJD
\INSTHA
\INSTID
\INSTIF
\INSTCC
\INSTCD
\INSTEJ
\INSTII
\INSTBA
\INSTFC
\INSTFI
\INSTIH
\INSTHB
\INSTCE
\INSTDF
\INSTFJ
\INSTEH
\INSTGJ
\INSTCF
\INSTGG
\INSTIC
\INSTGC
\INSTGD
\INSTHC
\INSTBC
\INSTJF
\INSTJB
\INSTFB
\INSTDI
\INSTIA
\INSTBB
\INSTJE
\INSTCH
\INSTBJ
\INSTCG
\INSTHF
\INSTGI
\INSTHG
\INSTB
\INSTJH
\INSTJJ
\INSTDJ
\INSTDH
\INSTIJ
\INSTFD
\INSTEA
\INSTHE
\INSTH

\author{K.\,Abe}\INSTBJ
\author{S.\,Abe}\INSTBJ
\author{H.\,Adhikary}\INSTDJ
\author{R.\,Akutsu}\INSTCB
\author{H.\,Alarakia-Charles}\INSTEJ
\author{Y.I.\,Alj Hakim}\INSTFB
\author{S.\,Alonso Monsalve}\INSTEF
\author{L.\,Anthony}\INSTEI
\author{S.\,Aoki}\INSTCC
\author{K.A.\,Apte}\INSTEI
% ...

\collaboration{The T2K Collaboration}\noaffiliation
    
\begin{abstract}
\noindent We report the first measurement of muon-neutrino charged-current cross section as a function of kinematic imbalance (KI) observables on oxygen with no pions and at least one proton in the final state, using the T2K ND280 detector. The cross section is extracted simultaneously for carbon and oxygen targets and double-differentially as a function of several KI observables, providing new insight into the modeling of nuclear effects. This joint measurement offers direct sensitivity to the correlations between two targets, a key ingredient for reducing systematic uncertainties in neutrino oscillation experiments that employ multiple target nuclei, such as T2K and Hyper-Kamiokande. Comparisons with predictions from widely used neutrino event generators show that none of the models fully describe the data across all regions of measured phase space. These results highlight possible directions where improvements in neutrino–nucleus interaction modeling are needed for current and future neutrino oscillation experiments.

\end{abstract}

\maketitle
{\em Introduction ---} Current and future long-baseline neutrino experiments, such as T2K~\cite{T2K:2011qtm} and Hyper-Kamiokande~\cite{Hyper-Kamiokande:2018ofw}, aim for precise measurements of neutrino oscillations. These analyses rely on comparing observed neutrino spectra to theoretical predictions, where at sub-GeV energies ($\sim$0.6~GeV), the signal is dominated by pionless 
charged-current (CC$0\pi$) interactions. Standard reconstruction of neutrino energy typically assumes quasi-elastic (QE) scattering off a free nucleon at rest~\cite{T2K:2023smv}. However, nuclear effects~\cite{Coloma:2013tba, Mosel:2016cwa, NuSTEC:2017hzk}, mainly including nucleon initial motion (Fermi motion), multi-nucleon correlations (short / long range correlations and n-particle n-hole interactions), and final-state interactions (FSIs), break this assumption. FSIs further complicate the topology by absorbing pions from pion production channels, migrating these events into the CC$0\pi$ sample. Mis-modeling these processes distorts neutrino energy reconstruction and introduces significant biases: current estimates suggest that alternative CC$0\pi$ models can bias the $\theta_{23}$, $\Delta m^{2}_{32}$, and $\delta_{CP}$ measurements by 25\%, 85\%, and 11\% of their systematic uncertainty budgets, respectively~\cite{T2K:2025yoy}. Refining neutrino-nucleus interaction models is therefore critical for the precise oscillation analyses.

\begin{figure}[htpb!]
\centering
\includegraphics[width=1.0\columnwidth, trim=0 10 0 25, clip]{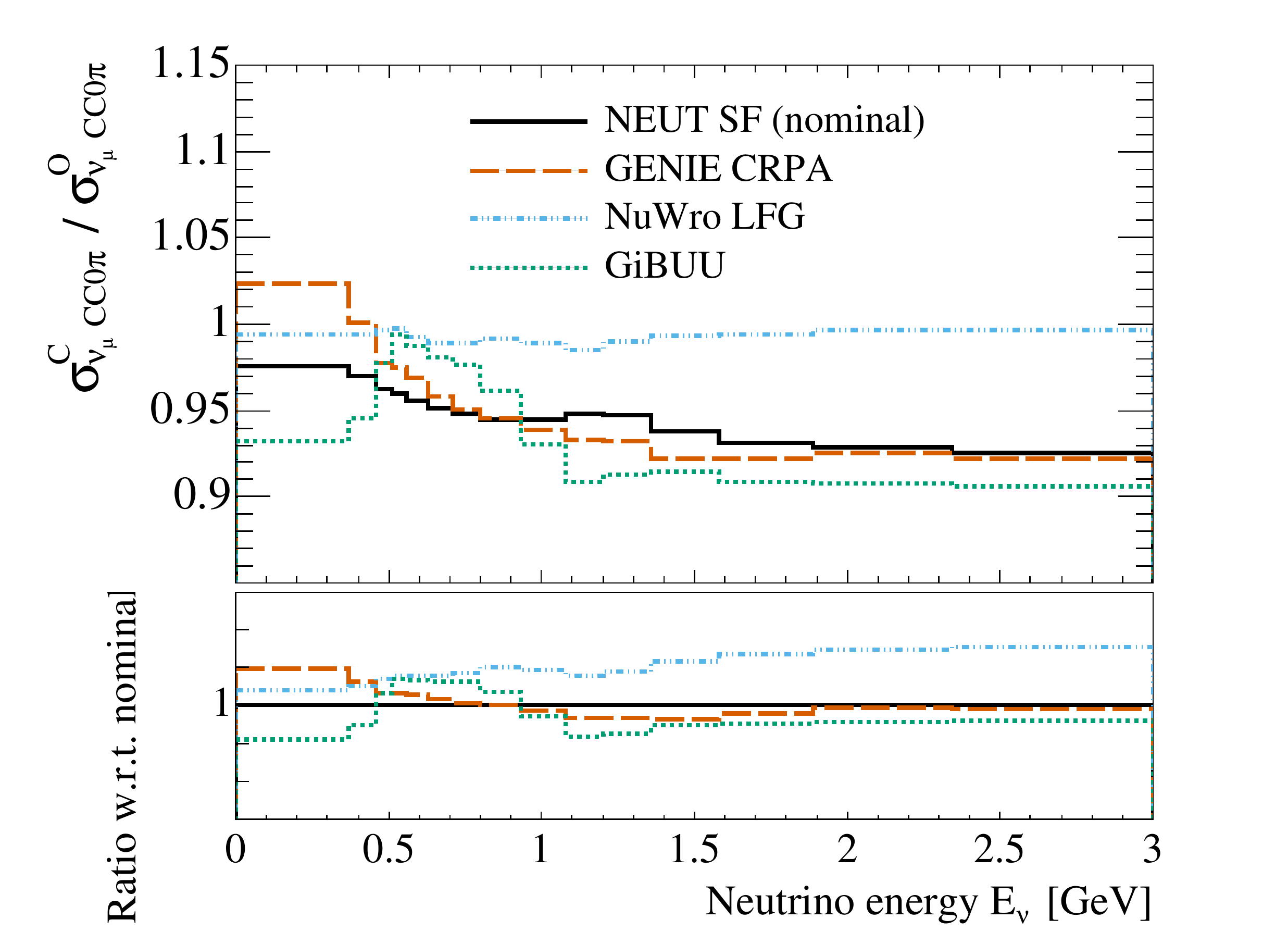}
\caption{Carbon-to-oxygen $\nu_{\mu}$ CC$0\pi$ cross-section ratios as a function of neutrino energy. (Top) Predicted ratios from four widely-used generators with distinct nuclear models: NEUT~\cite{Hayato:2009zz,Hayato:2021heg} with spectral function (SF) model~\cite{Benhar:1994hw}, GENIE~\cite{Andreopoulos:2009rq,GENIE:2021npt} with continuous random phase approximation (CRPA) model~\cite{Kolbe:1999au,Jachowicz:2002rr}, NuWro~\cite{Juszczak:2005zs,Golan2012nuwro} with local Fermi gas (LFG) model~\cite{Nieves:2011pp,Bourguille:2020bvw} and GiBUU~\cite{Buss:2011mx}. (Bottom) Ratios of each model relative to the nominal NEUT prediction. Discrepancies of up to $\sim$10\% highlight the model dependence inherent in extrapolating carbon-based measurements to oxygen targets.}
\label{fig:pred_COratio}
\end{figure}

While CC$0\pi$ interactions have been studied extensively~\cite{T2K:2020jav, T2K:2018rnz, T2K:2020sbd, MicroBooNE:2023cmw, MicroBooNE:2024yzp, MINERvA:2018hba}, measurements on oxygen remain scarce despite its important role as the primary target for Super-Kamiokande~\cite{Super-Kamiokande:2002weg} and Hyper-Kamiokande. Existing T2K oxygen measurements are limited to leptonic kinematics~\cite{T2K:2017qxv, T2K:2019ddy, T2K:2020jav}, which lack the sensitivity to hadronic final states needed to probe the complex nuclear effects. Besides, the T2K oscillation analyses largely rely on extrapolating carbon-based results from the ND280 near detector~\cite{T2K:2011qtm, T2K:2023smv} to oxygen. This extrapolation introduces significant model dependence. As shown in Fig.~\ref{fig:pred_COratio}, widely used interaction generators predict carbon-to-oxygen cross-section ratios differing by up to $\sim$10\%. Such discrepancies underline the necessity of direct measurements on multiple targets to resolve modeling uncertainties.

\begin{figure}[htpb!]
\centering
\includegraphics[width=0.35\textwidth]{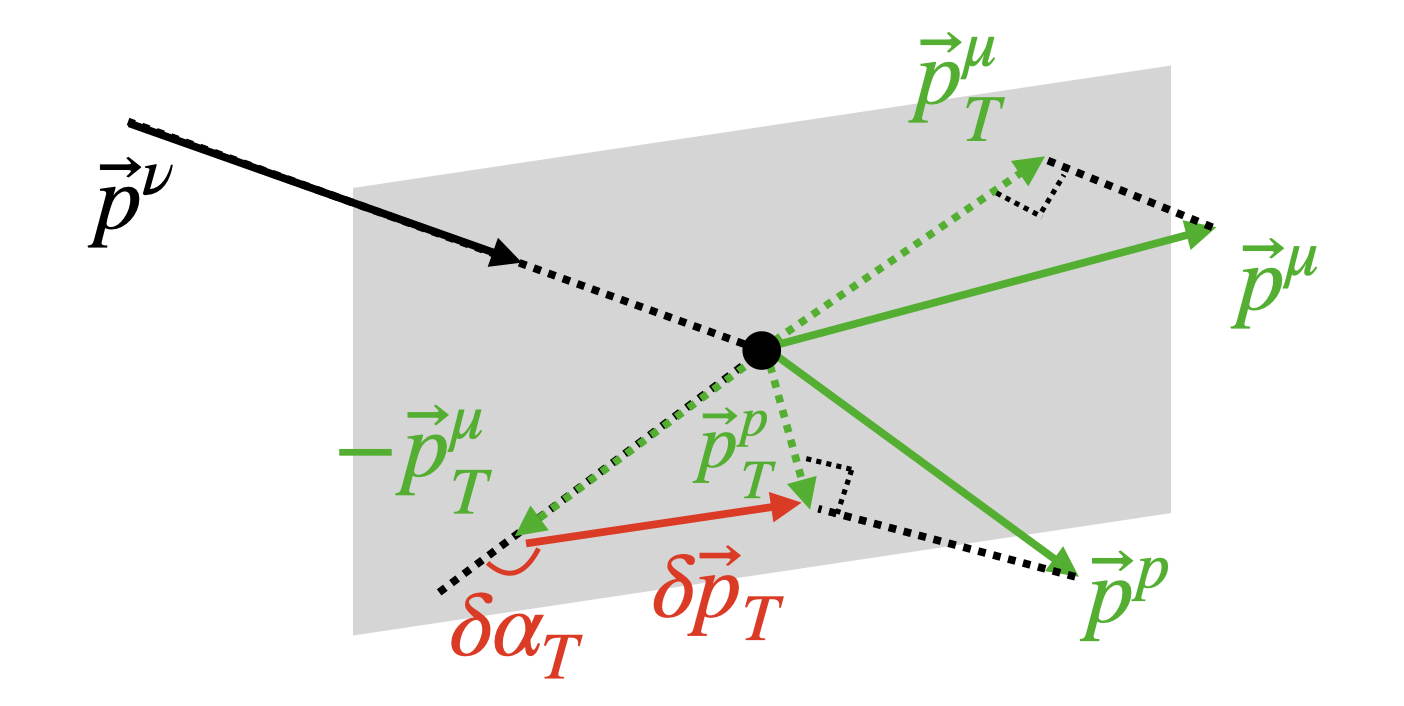}
\caption{A schematic depiction of the kinematic imbalance observables $\delta p_{T}$ and $\delta \alpha_{T}$ (shown in red). These quantities are defined on the plane transverse to the incident neutrino direction $\vec{p^{\nu}}$. The central black point indicates the target nucleus. The transverse momentum vectors of the outgoing muon and the leading proton are denoted by $\vec{p_{T}^{\mu}}$ and $\vec{p_{T}^{p}}$ (shown in green), respectively. The figure is adapted from Ref.~\cite{Lu:2015tcr}.}
\label{fig:TKI_sketch}
\end{figure}

To better resolve nuclear effects, recent measurements have increasingly utilized correlations between final-state leptons and hadrons~\cite{Lu:2015tcr, Lu:2019nmf}. Transverse kinematic imbalance (TKI) variables, already measured on carbon~\cite{T2K:2018rnz, MINERvA:2018hba} and argon~\cite{MicroBooNE:2023tzj}, exploit transverse momentum conservation to directly probe the initial nuclear state and FSI. More recently, ``generalized'' kinematic imbalance (GKI) observables have incorporated longitudinal information~\cite{MicroBooNE:2023krv, Furmanski:2016wqo}, providing further sensitivity but requiring an estimation of neutrino energy under QE assumption. For $\nu_\mu$ CC$0\pi$ channel, the KI observables are commonly reconstructed using the outgoing muon and the highest-momentum (leading) proton. In this Letter, we employ three KI observables: the transverse momentum imbalance $\delta p_{T}$~\cite{Lu:2015tcr}, the transverse boosting angle $\delta \alpha_{T}$~\cite{Lu:2015tcr}, and the inferred initial nucleon momentum $p_{N}$~\cite{Lu:2019nmf}. While $\delta p_{T}$ and $\delta \alpha_{T}$ are purely transverse (see Fig.~\ref{fig:TKI_sketch}), $p_{N} = \sqrt{\delta p_{T}^{2} + p_{L}^{2}}$ incorporates the longitudinal component $p_{L}$ derived from the energy-momentum conservation~\cite{Furmanski:2016wqo}, providing a comprehensive probe of the nuclear environment.

KI observables incorporate distinct contributions from a variety of nuclear effects. To disentangle the correlations of nuclear effects, we report double-differential cross sections for two complementary combinations: $\delta p_{T}$--$\delta \alpha_{T}$ and $p_{N}$--$\cos\theta_{\mu}$, where $\cos\theta_{\mu}$ denotes the muon scattering angle relative to the incident neutrino direction. The $\delta p_{T}$--$\delta \alpha_{T}$ space effectively separates initial-state dynamics from FSI. While $\delta p_{T}$ reflects the transverse nucleon momentum, FSI enhances the high-$\delta p_{T}$ tail through hadron deceleration. Similarly, FSI shifts the $\delta \alpha_{T}$ distribution toward higher values, leaving the low-$\delta \alpha_{T}$ region dominated by Fermi motion. This combination thus isolates FSI-dominated events from the initial-state bulk. In the $p_{N}$--$\cos\theta_{\mu}$ space, the muon scattering angle distinguishes interaction channels. CCQE processes dominate at large angles, while forward scattering receives substantial contributions from multi-nucleon correlations and pion-absorbing FSIs. The CCQE contribution in the forward-scattering region is also suppressed, because the relatively high proton selection threshold applied in this analysis (described later) reduces the low-energy-transfer contribution. Because $p_{N}$ incorporates longitudinal information, its bulk tracks the initial nucleon motion while its tail remains sensitive to non-QE processes and FSI. Together, these 2D observables provide a robust framework for isolating CCQE-dominated interactions from non-QE processes and FSI.

In this Letter, we present the first double-differential cross-section measurements of pionless muon-neutrino charged-current scattering with final-state protons identified ($\nu_{\mu}$ CC$0\pi Np$) on both carbon and oxygen as functions of $\delta p_{T}$--$\delta \alpha_{T}$ and $p_{N}$--$\cos\theta_{\mu}$. In addition to probing the nuclear effects, these results provide new insights into the joint description of carbon and oxygen neutrino–nucleus scattering which is directly relevant to the physics programs of T2K and Hyper-Kamiokande. Furthermore, the measurements on oxygen offer a unique opportunity to probe nuclear effects in a ``doubly magic’’ nucleus, in which both proton and neutron shells are filled.

{\em Experimental setup ---} Tokai-to-Kamioka (T2K) experiment~\cite{T2K:2011qtm} measures neutrino oscillations using a high-purity neutrino beam produced at the J-PARC facility. The beam is characterized prior to oscillations by the ND280 near detector, located 280~m downstream of the production target. The oscillated spectrum is then measured 295~km away at the Super-Kamiokande far detector. Both detectors are positioned at an off-axis angle of $2.5^{\circ}$ relative to the beam direction and see the neutrino energy spectrum that peaks at 0.6 GeV~\cite{T2K:2012bge}. This configuration produces a narrow-band neutrino energy spectrum and maximizes the oscillation probability at the far detector~\cite{E899:1995bzq}.

This measurement uses the neutrino beam data collected with the ND280, a magnetized detector composed of several subdetector systems. The analysis employs the two fine-grained scintillator detectors (FGDs)~\cite{T2KND280FGD:2012umz} as neutrino interaction targets. FGD1 consists of layers of plastic scintillator and serves as a carbon target, while FGD2 is composed of interleaved plastic scintillator and water layers, providing both carbon and oxygen targets. The FGDs are sandwiched between three time projection chambers (TPCs)~\cite{T2KND280TPC:2010nnd}, which track and measure charged particles produced from neutrino interactions within the active targets. Surrounding electromagnetic calorimeters (ECALs)~\cite{T2KUK:2013wkh} measure the energy deposited by particles exiting the inner detectors.

{\em Analysis samples ---} The signal considered in this measurement consists of $\nu_{\mu}$ CC$0\pi Np$ interactions occurring on carbon or oxygen within the fiducial volumes of the ND280 FGD1 and FGD2 detectors. FGD1 samples are dominated by interactions on carbon, whereas FGD2 samples consist of a mixture of carbon and oxygen targets. In addition, phase-space constraints are applied to the selected muon and leading proton to restrict the measurement to kinematic regions with good detector acceptance and well-understood detector systematic uncertainties. These constraints are defined as $0.225 < p_{\mu} < 10$~GeV/$c$ and $\cos\theta_{\mu} > -0.6$ for the muon momentum and scattering angle, and $0.525 < p_{p} < 1$~GeV/$c$ and $\cos\theta_{p} > 0.3$ for the leading proton momentum and scattering angle. The applied phase-space constraint has no significant impact on the neutrino energy coverage of the analysis samples in the flux peak region at $\sim$ 0.6~GeV.

This measurement uses T2K neutrino-mode data collected between 2010 and 2017, corresponding to a total exposure of $11.61 \times 10^{20}$ protons on target (POT). A total of 9469 $\nu_{\mu}$ CC$0\pi Np$ signal candidates are selected, with an expected composition of 70.8\% interactions on carbon and 22.3\% on oxygen~\footnote{The remaining 6.9\% of interactions occurred on other materials present in the detector, namely titanium, silicon and nitrogen, composing the glue and optical fibers~\cite{T2KND280FGD:2012umz}.}. The overall signal selection purity is 74.6\%, with an efficiency of 25.6\%. The efficiency is primarily limited by the proton tagging capabilities of the FGDs and TPCs, specifically the reconstruction and selection requirements for the leading proton. For the carbon interactions, the signal selection purity and efficiency are 81.3\% and 26.5\%, respectively, while for oxygen they are 76.4\% and 22.8\%. The reduced performance for oxygen is primarily due to the structure of FGD2, which contains inactive water layers, making the reconstruction of low-momentum protons more challenging~\cite{T2KND280FGD:2012umz}.

The dominant background arises from single-pion production ($\nu_{\mu}$~CC$1\pi^{+}$) events where the outgoing $\pi^{+}$ escapes detection in the FGDs or TPCs, either due to $\pi^{+}$ momentum below the reconstruction threshold or $\pi^{+}$ undergoing secondary hadronic interactions. Consequently, these events are misidentified as the $\nu_{\mu}$~CC$0\pi$ signal. The $\nu_{\mu}$~CC$1\pi^{+}$ background constitutes approximately 9.4\% of the selected sample. To accurately estimate and constrain this contribution, a dedicated control sample enriched in $\nu_{\mu}$ CC$1\pi^{+}$ interactions is developed, and a data-driven method is applied to constrain and subtract this background from the signal sample. Ref.~\cite{Abe:2026bjg} summarizes the full selection details and performances.

{\em Analysis strategy ---} This measurement extracts the $\nu_{\mu}$ CC$0\pi Np$ cross section using a template-fit unfolding method~\cite{T2K:2025kdk, T2K:2025kda, T2K:2023qjb} with the GUNDAM software~\cite{GUNDAM}. The fit model is constructed from selected Monte Carlo (MC) samples and includes free normalization parameters to account for discrepancies between data and MC predictions. The fit model is subject to several sources of systematic uncertainty, including those associated with detector simulation, neutrino flux modeling, neutrino–nucleus interaction modeling for both signal and background processes, and number of target nucleons. Among these, the interaction modeling uncertainties constitute the dominant contribution to the total systematic error. To validate the cross-section extraction framework, fits to pseudo-data generated with alternative neutrino–nucleus interaction models were performed. In all cases, the extracted cross sections were consistent with the input pseudo-data, demonstrating the robustness of the analysis method. A detailed description of the template-fit approach, as well as a summary of the systematic uncertainties and validation studies, is provided in Ref.~\cite{Abe:2026bjg}.

{\em Comparison to generator predictions ---} The measured cross sections are compared with predictions from several neutrino event generators using the NUISANCE framework~\cite{Stowell:2016jfr}. The predictions are obtained from NEUT v6.1.0~\cite{Hayato:2009zz,Hayato:2021heg}, GENIE v3.6.2~\cite{Andreopoulos:2009rq,GENIE:2021npt}, NuWro v25.03.1~\cite{Juszczak:2005zs}, and GiBUU v2025p3~\cite{Buss:2011mx}. Together, these configurations span a range of models describing nucleon initial-state dynamics, multi-nucleon interactions, and FSI. Further comparisons focusing on variations of the models that describe specific interaction modes can be found in Ref.~\cite{Abe:2026bjg}.

To model the nucleon initial state, we consider the spectral function (SF) approach~\cite{Benhar:1994hw}, implemented in NEUT (with an updated proton SF for carbon~\cite{Ankowski:2024ntv}) and NuWro, as well as the local Fermi gas (LFG) model~\cite{Nieves:2011pp,Bourguille:2020bvw} used in GENIE. The GENIE predictions are generated using two comprehensive model configurations (CMCs), G18\_10b\_02\_11b and AR23\_20i\_00\_000 (hereafter referred to as G18\_10b'' and AR23\_20i’’). The AR23\_20i configuration extends the initial nucleon momentum distribution beyond the Fermi momentum ($p_{N} \gtrsim 0.3$GeV/$c$) in CCQE interactions to account for short-range correlated (SRC) pairs~\cite{Filali:2024vpy}. GiBUU models the nucleon initial state by placing the nucleons in a momentum-dependent mean-field potential, with their momentum distribution described by the LFG model~\cite{Gallmeister:2016dnq}.

For multi-nucleon interactions, we consider the Valencia model~\cite{Nieves:2011pp,Gran:2013kda,Sobczyk:2020dkn} to describe two-particle two-hole (2p2h) interactions, as implemented in NEUT, NuWro, and GENIE G18\_10b, along with the SuSAv2 model~\cite{Gonzalez-Jimenez:2014eqa,Dolan:2019bxf} utilized in GENIE AR23\_20i. The NuWro implementation further provides exclusive predictions for the kinematics of the final-state nucleons~\cite{Prasad:2024gnv,Sobczyk:2024ecl}. Additionally, three-particle three-hole (3p3h) contributions are included in both NEUT and NuWro, following an implementation~\cite{Prasad:2024gnv} based on an extension of the Valencia model. In GiBUU, 2p2h interactions are described by a phenomenological model guided by electron-scattering data~\cite{Gallmeister:2016dnq,Bodek:2022gli}.

Resonant pion production is modeled based on the Rein–Sehgal model~\cite{Rein:1980wg} in NEUT, GENIE and NuWro, except for GiBUU which has its own treatments~\cite{Mosel:2017nzk,Lalakulich:2010ss,Drechsel:2007if}. More inelastic channels (for example, deep inelastic interactions~\cite{Bodek:2005de}) are treated differently by each generator. However, they are not discussed in detail here, as they contribute only 2.4\% and 2.2\% to the carbon and oxygen signal cross sections extracted by this measurement, respectively.

Finally, the FSIs are modeled using the intranuclear cascade (INC) approaches~\cite{Serber:1947zza,Salcedo:1987md}, implemented with different variations in NEUT~\cite{Hayato:2021heg}, NuWro~\cite{Golan2012nuwro}, and GENIE~\cite{Dytman:2021ohr}. In GENIE, the G18\_10b and AR23\_20i CMCs employ the hN2018 and hA2018 FSI models, respectively. The hN2018 model follows a multi-step INC approach, while hA2018 is an empirical model that simulates FSI through a single effective interaction. GiBUU adopts a different strategy for hadron transport, based on a ``time-like'' rather than a ``space-like'' approach by solving the Boltzmann–Uehling–Uhlenbeck equation~\cite{Mosel:2016cwa,Mosel:2019vhx}.

\begin{figure*}
\centering
\subfloat[$\delta p_{T}$--$\delta \alpha_{T}$]
{\includegraphics[width=1.0\textwidth]{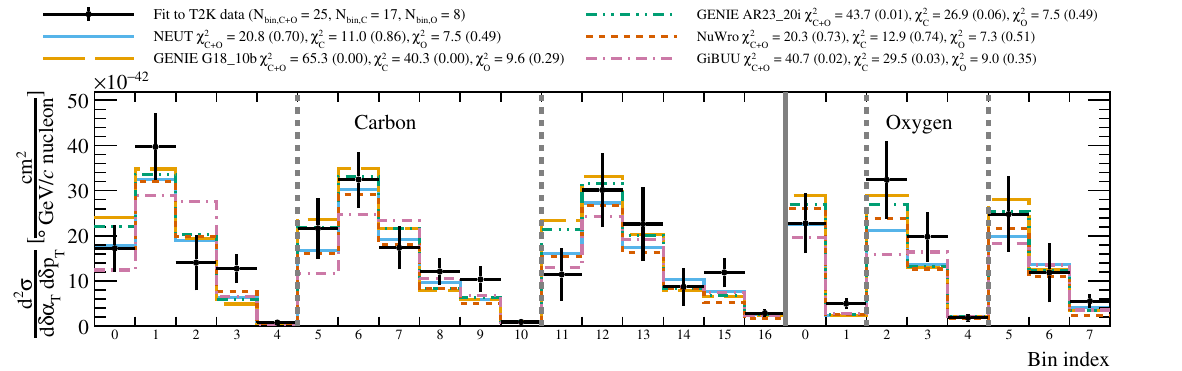}}
\\
\subfloat[$p_{N}$--$\cos\theta_{\mu}$]
{\includegraphics[width=1.0\textwidth]{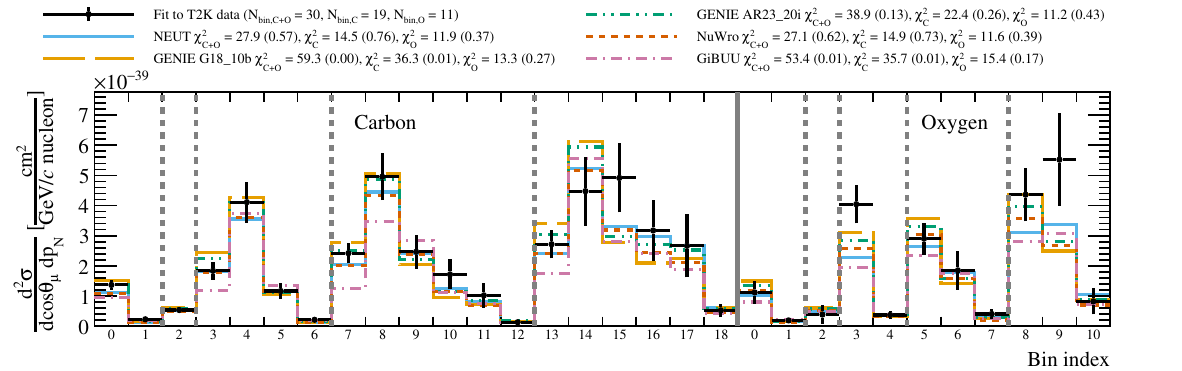}}
\caption{Measurements of the $\nu_{\mu}$ CC$0\pi Np$ double-differential cross sections jointly on carbon and oxygen targets, obtained from the fit to T2K data in the (a) $\delta p_{T}$--$\delta \alpha_{T}$ and (b) $p_{N}$--$\cos\theta_{\mu}$ phase spaces. Error bars represent the combined statistical and systematic uncertainties. The measurements are compared with various neutrino-nucleus interaction models: NEUT (sky blue), GENIE G18\_10b (orange), GENIE AR23\_20i (bluish green), NuWro (vermillion), and GiBUU (reddish purple). The $\chi^{2}$ and corresponding $p$-values (in parentheses) are reported for the joint carbon-oxygen dataset ($\chi^{2}_{C+O}$), as well as for the carbon ($\chi^{2}_{C}$) and oxygen ($\chi^{2}_{O}$) components individually. The vertical gray solid lines separate carbon and oxygen results and dashed lines separate $\delta \alpha_{T}$ (a) and $\cos\theta_{\mu}$ (b) slices as illustrated in Fig.~\ref{fig:xsec_binning}. Bin indices are defined according to the schemes shown in Fig.~\ref{fig:xsec_binning}.}
\label{fig:modelcomp_plot_PRL}
\end{figure*}

\begin{figure}
\centering
\subfloat[$\delta p_{T}$--$\delta \alpha_{T}$]
{\includegraphics[width=0.45\textwidth]{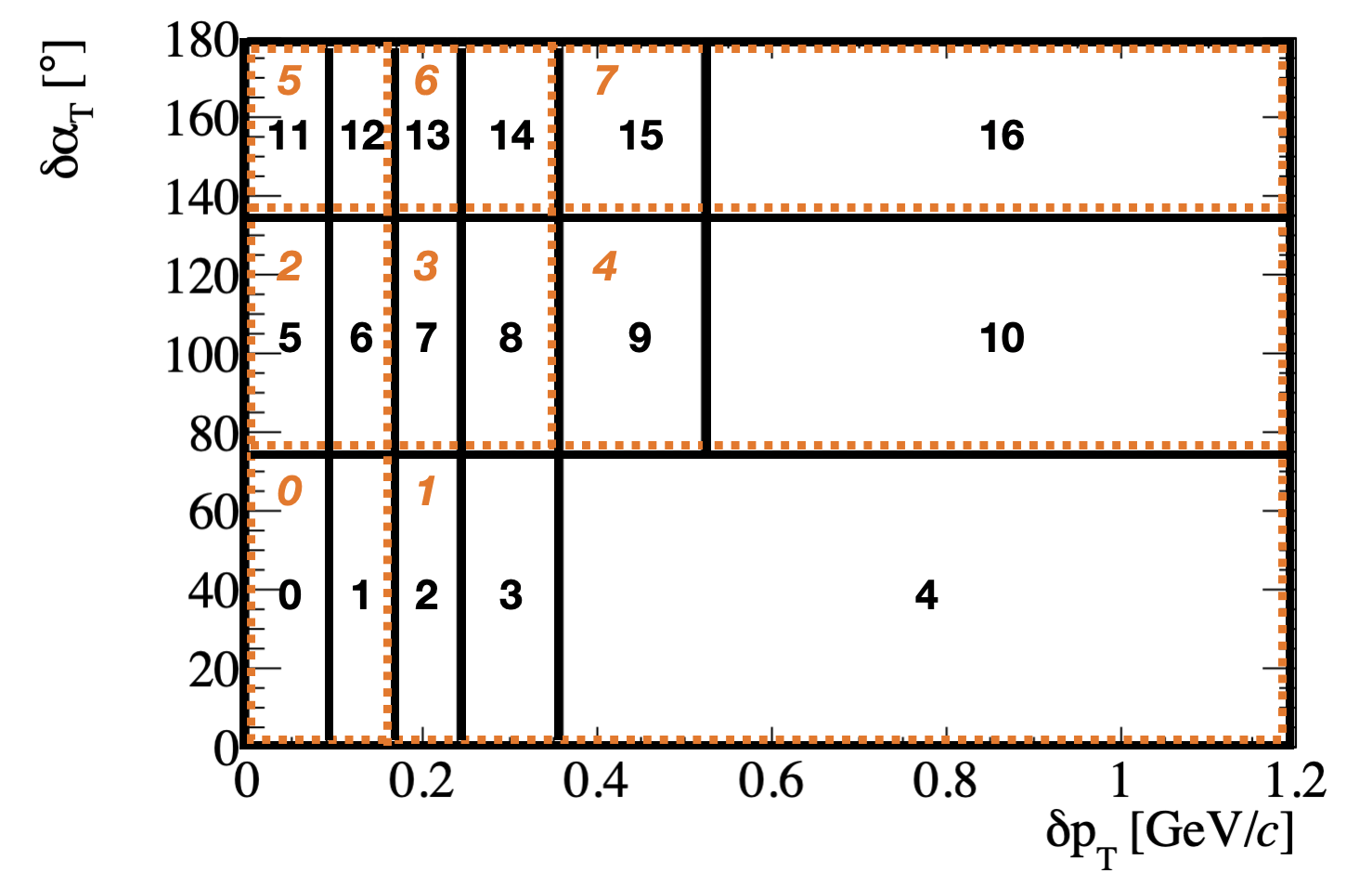}}
\\
\subfloat[$p_{N}$--$\cos\theta_{\mu}$]
{\includegraphics[width=0.45\textwidth]{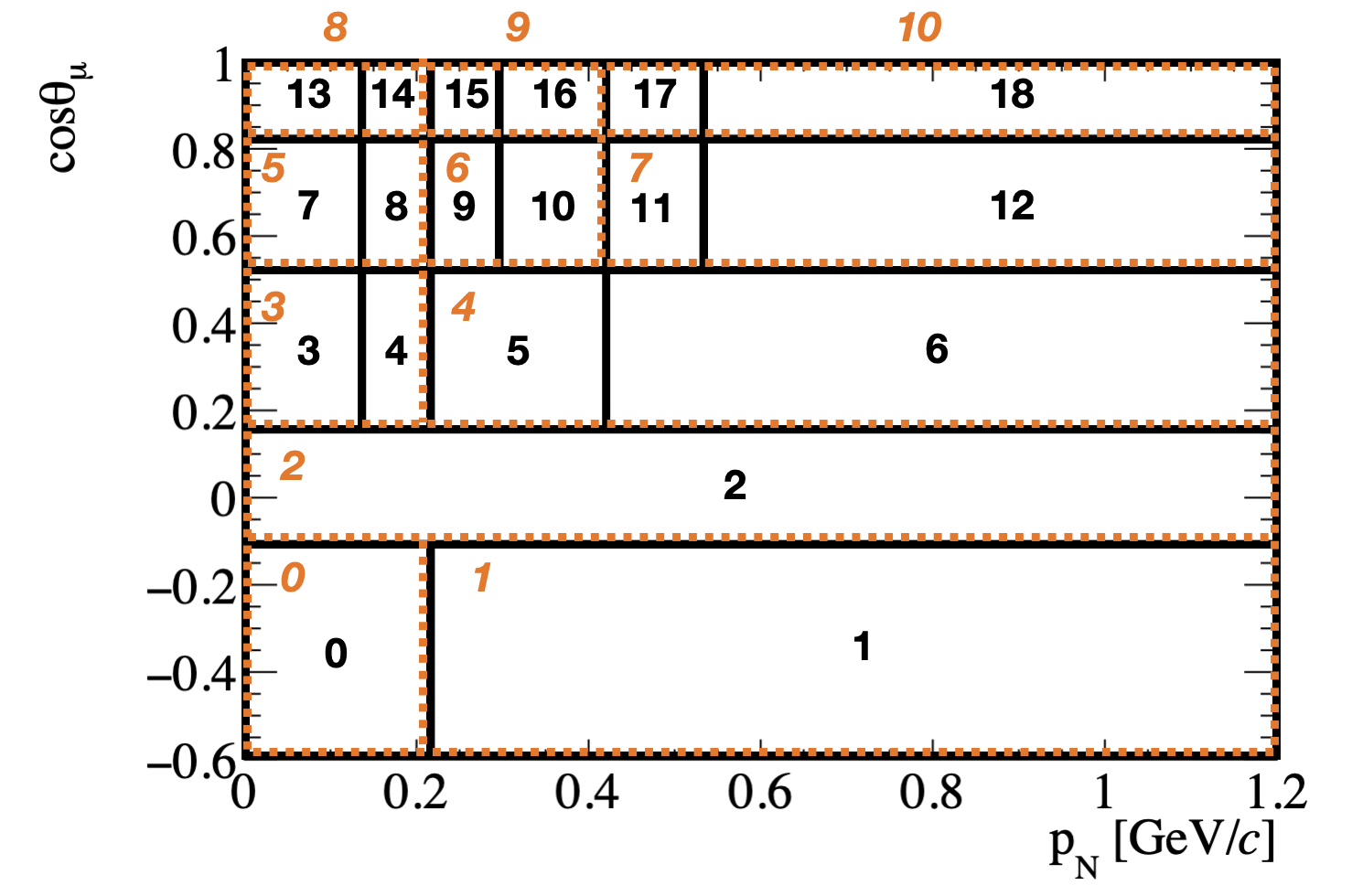}}
\caption{Binning schemes utilized for the cross-section measurements in the (a) $\delta p_{T}$--$\delta \alpha_{T}$ and (b) $p_{N}$--$\cos\theta_{\mu}$ phase spaces. Regions delineated by solid black lines represent the binning applied to the carbon measurement, while regions enclosed by dashed orange lines indicate the binning for the oxygen measurement. The bin indices shown here correspond to the labeling used in Fig.~\ref{fig:modelcomp_plot_PRL}.}
\label{fig:xsec_binning}
\end{figure}

{\em Results ---} Fig.~\ref{fig:modelcomp_plot_PRL} presents the extracted cross sections compared with generator predictions in the $\delta p_{T}$--$\delta \alpha_{T}$ and $p_{N}$--$\cos\theta_{\mu}$ phase spaces, respectively. The corresponding binning schemes applied in this measurement are illustrated in Fig.~\ref{fig:xsec_binning}. The measured cross sections exhibit total relative uncertainties ranging from 15\% to 80\% across the analysis bins. The total uncertainty is dominated by statistical fluctuations, which account for 9\%--70\% of the error depending on the bin. Among the systematic sources, neutrino-nucleus interaction modeling is the primary contribution, ranging from 8\% to 30\%. This is followed by detector simulation and neutrino flux modeling, which provide comparable contributions of 2\%--6\%. The uncertainty associated with the number of target nucleons remains below 1\% across all bins. To assess the level of agreement between the measurements and model predictions, $\chi^{2}$ values and the corresponding $p$-values are computed for each comparison. The $\chi^{2}$ is defined as
\begin{equation}
    \chi^{2} = \sum_{i,j} \Delta^{\sigma}_{i} (V^{-1})_{ij} \Delta^{\sigma}_{j},
\end{equation}
where the sum goes through all differential cross-section bins (denoted as $i$ and $j$), $\Delta^{\sigma}_{i}$ is the difference between the predicted and measured cross section in bin $i$ (same for bin $j$) and $V$ is the covariance matrix describing the uncertainties and correlations of the measured cross section across different bins. The uncertainties are assumed to be Gaussian distributed. The $p$-values are evaluated under the assumption that the $\chi^{2}$ follows the expected probability distribution with the number of degrees of freedom equals to the number of bins used in the cross-section extraction. A small $p$-value, defined here as smaller than or equal to 0.05, indicates that the model provides a poor description of the data. It should be noted, however, that the theoretical uncertainties on the model predictions are not included in the $\chi^{2}$ evaluation, as they are not available from all generators. The $\chi^{2}$ values thus reflect the choice of central parameter values in the models considered.

{\em Discussion of the results ---} Overall, the NEUT and NuWro predictions show the best agreement with both carbon and oxygen data, yielding the highest $p$-values across all measured phase spaces. These results suggest a preference for the SF ground state, the Valencia 2p2h model, and the NEUT/NuWro-style INC for FSI. In contrast, GENIE and GiBUU exhibit larger tension, particularly in the carbon part, though their oxygen predictions remain more consistent with the data. Notably, the GENIE AR23\_20i configuration achieves a better $p$-value ($\sim$0.13) in the $p_{N}$--$\cos\theta_{\mu}$ space. Compared to another GENIE prediction, G18\_10b, the AR23\_20i CMC incorporates a larger contribution from SRCs in modeling CCQE interactions and employs the SuSAv2 model for 2p2h processes, which potentially result in an improved description of the data.

Examining the results across different $\delta \alpha_{T}$ regions, we find that GiBUU accurately describes the high-angle region ($\delta \alpha_{T} > 135^{\circ}$), where FSI effects are most pronounced. However, the model fails to capture the low-$\delta \alpha_{T}$ region, where CCQE interactions with weaker FSI dominate, due to an underestimated $\delta p_{T}$ bulk. A consistent trend appears in the $p_{N}$--$\cos\theta_{\mu}$ space: GiBUU provides a reasonable description for forward-going muons ($\cos\theta_{\mu} > 0.83$), yet diverges at larger scattering angles where the nucleon ground-state modeling of CCQE becomes the primary driver of the $p_{N}$ peak. These discrepancies suggest that GiBUU may either overestimate FSI strengths or underestimate the CCQE contribution within these measured phase spaces.

Despite varying modeling approaches, all tested generators under-predict the data in the intermediate $\delta p_{T}$ region ($0.2$--$0.5$~GeV/$c$) for carbon. A similar, though less pronounced, discrepancy persists for oxygen at intermediate $75^{\circ} < \delta\alpha_{T} \leqslant 135^{\circ}$ region. A comparable deficit appears in the intermediate $p_{N}$ range ($0.2$--$0.4$~GeV/$c$) for forward-going muons ($\cos\theta_{\mu} > 0.83$) across both targets. These kinematic regions are sensitive to SRCs and multi-nucleon interactions such as npnh, suggesting that current implementations of these processes might be insufficient. Furthermore, for the oxygen target, the data exhibit a higher cross section near the $p_{N}$ peak at large muon scattering angles ($0.15 < \cos\theta_{\mu} < 0.52$). This deviation potentially underlines the need for a more refined treatment of the initial nuclear state, particularly for oxygen CCQE interactions, in this phase space region.

{\em Conclusions ---} We present the first joint double-differential $\nu_{\mu}$ CC$0\pi Np$ cross-section measurements on carbon and oxygen, performed in the $\delta p_{T}$--$\delta \alpha_{T}$ and $p_{N}$--$\cos\theta_{\mu}$ phase spaces. These results provide a novel and highly sensitive probe of nuclear effects, including nucleon initial-state motion, multi-nucleon interactions, and FSIs. Furthermore, this work offers a direct test of the unified modeling of neutrino-nucleus scattering on carbon and oxygen relevant for T2K and Hyper-Kamiokande. This measurement is complementary to the multi-target studies at higher neutrino energies reported in Ref.~\cite{MINERvA:2025tem}.

Comparisons with a broad range of interaction models show that generators implementing a SF ground state, Valencia model for multi-nucleon correlations, and cascade-based FSI from NEUT and NuWro, provide the best overall description of the carbon and oxygen data. Nevertheless, notable discrepancies remain, particularly in the intermediate $\delta p_{T}$ and $p_{N}$ regions, where all models under-predict the measurements for both targets. This points to missing or underestimated contributions, potentially associated with the description of SRCs or multi-nucleon interactions.

The measurement remains limited by the statistical uncertainties. Future T2K data collected with the upgraded ND280 detector~\cite{T2K:2019bbb}, including the installation of the new Super-FGD neutrino target~\cite{Abe:2026elv}, will provide substantially larger event samples and improved reconstruction of low-momentum protons, thereby enhancing sensitivity to the KI observables. These developments, together with the methodology and insights established in this first measurement, will enable more precise studies of nuclear effects and help to reduce cross-section uncertainties in future oscillation analyses.

{\em Acknowledgments ---} The T2K collaboration would like to thank the J-PARC staff for superb accelerator performance. We thank the CERN NA61/SHINE Collaboration for providing valuable particle production data. We acknowledge the support of MEXT,   JSPS KAKENHI  and bilateral programs, Japan; UGent-BOF and FWO-Flanders, Belgium; NSERC, the NRC, and CFI, Canada; the CEA and CNRS/IN2P3, France; the Deutsche Forschungsgemeinschaft (DFG 397763730, 517206441), Germany; the NKFIH  (NKFIH 137812 and TKP2021-NKTA-64), Hungary; the INFN, Italy; the Ministry of Science and Higher Education (2023/WK/04) and the National Science Centre (UMO-2018/30/E/ST2/00441, UMO-2022/46/E/ST2/00336 and UMO-2021/43/D/ST2/01504), Poland; the RSF (RSF 26-12-00495) and the Ministry of Science and Higher Education, Russia;  MICINN  (PID2022-136297NB-I00 /AEI/10.13039/501100011033/ FEDER, UE, PID2024-157541NB-I00 (UAM) and PID2023-146401NB-I00 (US), Severo Ochoa Centres of Excellence Programme 2025-2029(CEX2024001441-S),  Government of Andalucia (FQM160) and the University of Tokyo ICRR's Inter-University Research Program FY2025 Ref. J1, and ERDF and European Union (UAM: H2020-MSCA-RISE-GA872549- SK2HK) and NextGenerationEU funds (PRTR-C17.I1) and  Generalitat de Catalunya (AGAUR 2021-SGR-01506, CERCA program) University of Seville grant (RYC2022-035203-I funded by MICIU/AEI/10.13039/501100011033, ``ERDF a way of making Europe'' and FSE+, Ayudas ``Atracción de Investigadores con Alto Potencial''. Ref. VIIPPIT-2025, and Secretariat for Universities and Research of the Ministry of Business and Knowledge of the Government of Catalonia and the European Social Fund (2022FI\_B 00336), Spain; the SNSF and SERI, Switzerland; the STFC and UKRI, UK; the DOE, USA; and NAFOSTED (103.99-2023.144,IZVSZ2.203433), Vietnam. We also thank CERN for the UA1/NOMAD magnet, DESY for the HERA-B magnet mover system, the BC DRI Group, Prairie DRI Group, ACENET, SciNet, and CalculQuebec consortia in the Digital Research Alliance of Canada, and GridPP in the United Kingdom, the CNRS/IN2P3 Computing Center in France and NERSC, USA. In addition, the participation of individual researchers and institutions has been further supported by funds from the ERC (FP7), “la Caixa” Foundation, the European Union’s Horizon 2020 Research and Innovation Programme under the Marie Sklodowska-Curie grant, the Horizon Europe Marie Sklodowska-Curie Staff Exchange project JENNIFER3 grant 101183137; the JSPS, Japan; the Royal Society, UK; French ANR and Sorbonne Université Emergences programmes; the VAST-JSPS (No. QTJP01.02/20-22);  and the DOE Early Career programme, USA. For the purposes of open access, the authors have applied a Creative Commons Attribution license to any Author Accepted Manuscript version arising.

{\em Data availability ---} The data that support the findings of this article are openly available~\cite{data_release}.

\bibliographystyle{apsrev4-2}
\bibliography{biblio}

\end{document}